\documentclass[twocolumn, floatfix, amsmath, showpacs]{revtex4}
\usepackage{graphicx}% Include figure files

\begin{document}

\title{Spatial homogeneity of optically switched semiconductor photonic
crystals and of bulk semiconductors}

\author{Tijmen G. Euser}\email{T.G.Euser@utwente.nl}
\author{Willem L. Vos}
\affiliation{Complex Photonic Systems (COPS), Department of
Science and Technology and MESA$^+$ Research Institute, University
of Twente, P.O. Box 217, 7500 AE Enschede, The Netherlands}
\homepage{www.photonicbandgaps.com}


\begin{abstract}

This paper discusses free carrier generation by pulsed laser
fields as a mechanism to switch the optical properties of
semiconductor photonic crystals and bulk semiconductors on an
ultrafast time scale. Requirements are set for the switching
magnitude, the time-scale, the induced absorption as well as the
spatial homogeneity, in particular for silicon at
$\lambda$~=~1550~nm. Using a nonlinear absorption model, we
calculate carrier depth profiles and define a homogeneity length
$\ell_{\rm{hom}}$. Homogeneity length contours are visualized in a
plane spanned by the linear and two-photon absorption
coefficients. Such a generalized homogeneity plot allows us to
find optimum switching conditions at pump frequencies near
$\nu$/c=5000~cm$^{-1}$ ($\lambda$~=~2000~nm). We discuss the
effect of scattering in photonic crystals on the homogeneity. We
experimentally demonstrate a $10\%$ refractive index switch in
bulk silicon within $230$~fs with a lateral homogeneity of more
than 30~$\rm{\mu}$m. Our results are relevant for switching of
modulators in absence of photonic crystals.
\end{abstract}

\maketitle

\newpage
\section{Introduction}

There is a fast growing interest in photonic crystals; composite
materials whose refractive index varies periodically on length
scales that match the wavelength of light.\cite{Crete} The optical
properties of photonic materials are determined by the spatially
varying refractive index, analogous to the periodic potential for
an electron in a crystal. Large spatial variations of the
refractive index cause a strong interaction between light and the
composite structure. Bragg diffraction causes the photonic
dispersion to organize into bands, much like the energy levels of
electrons in semiconductors. A major goal of the field is the
realization of three-dimensional (3D) structures that possess a
photonic band gap.\cite{Yablonovitch87,John87} At frequencies
inside the band gap, the optical density of states vanishes. This
should completely inhibit spontaneous emission of sources inside
the photonic crystal.\cite{Yablonovitch87} Indeed, strong
modifications of the spontaneous emission lifetime of quantum dots
have recently been demonstrated in photonic
crystals.\cite{lodahl04} In the presence of weak controlled
disorder, Anderson localization of light is also
predicted.\cite{John87} In this case, a photon may be trapped at a
point defect which serves as a cavity with a high quality
factor.\cite{Yablonovitch91}

In the examples above, the photonic crystals themselves do not
change in time. Switching experiments, in which the properties of
photonic crystals are modified on an ultrafast timescale allow
many interesting new opportunities. Switching 3D photonic crystals
is particularly interesting, as it provides the dynamic control
over the density of states inside the crystal as well as a change
in Bragg reflections.\cite{Johnson02} Ultrafast control of the
density of states should allow the switching of spontaneous
emission of light sources inside a crystal, and capturing and
releasing light in cavities inside the crystal. In absence of
photonic crystals, ultrafast switching of bulk semiconductors
finds applications in high speed optical modulators\cite{intel04}
and waveguides.\cite{Bristow03} Last but not least, ultrafast
control of photonic crystals is important for controlling the
propagation of light, such as in switched macroporous
silicon,\cite{Leonard02} or 2D crystal slabs.\cite{Cleo04}

In optical switching experiments, four important requirements have
to be met.\cite{Johnson02} First of all, the magnitude of the
induced change in the real part of the refractive index $n'$ must
be large enough to obtain the desired effect. A relative change in
$n'$ of $5\%$ is required to induce a major change in the density
of states. Such large changes can be induced by free carrier
generation\cite{Sokolowski00} but are not achievable with Kerr
switching. The second requirement, which is important for
applications, is the minimum time scale $\Delta$t over which the
switch occurs. In experiments where light pulses are trapped
inside photonic crystals, a switching time scale on the order of a
few hundred femtoseconds is necessary. The third requirement is
that the absorption of probe light, gauged by $n''$, should be
small in the switched sample. Excited carriers in a semiconductor
bring about inevitable absorption (related by a Kramers-Kronig
relation to the change in $n'$), that should remain within limits
by a limited carrier density. The fourth requirement concerns the
spatial homogeneity of the change $\Delta n'$ in a sample.
Homogeneity is particularly important in switching the density of
states in 3D photonic crystals.  A large gradient in $\Delta n'$
in the crystal results in a highly chirped switched sample, which
can no longer be considered a photonic crystal. All of these
requirements also pertain to other applications of switched
semiconductors, such as in waveguiding,\cite{intel04} albeit much
relaxed. Therefore, we expect the results from the present study
also to be relevant for applications outside photonic crystals.

Switching experiments in photonic crystals were pioneered by
Leonard \emph{et al.}.\cite{Leonard02} Optical free carrier
generation was used to change the refractive index of the backbone
of a 2D silicon photonic crystal. The carrier density generated in
their experiment was sufficiently high to induce a shift of a
Bragg stopgap on ultrafast time scales. Nevertheless, the
experiments revealed a serious limitation: the absorption of the
pump beam limits the volume of switched material and leads to a
spatial inhomogeneity in the degree of switching. Leonard \emph{et
al.} deduced that only a layer of three unit cells near the sample
surface was switched.\cite{Leonard02} Similar inhomogeneity is
probably also playing a role in recent studies of silicon
infiltrated opaline crystals, where a disappearance of the Bragg
peak was observed in the absorption range.\cite{Mazur03}
Therefore, the use of two-photon absorption was proposed as a way
to increase the penetration depth of pump light into the sample
and improve the switching homogeneity.\cite{Johnson02}

In this paper, we investigate the spatial homogeneity of optically
generated free carrier plasmas in semiconductors. We discuss a
non-linear absorption model that takes into account both linear
and two-photon absorption processes. From this model, we derive
optimum pumping conditions and we define a homogeneity length
scale to obtain homogeneous switching conditions. In particular
we trace constant homogeneity lengths in a general
parameter-diagram that pertains to any semiconductor. We discuss
the role of disorder-induced diffusion of the pump beam in
photonic crystals. Finally we experimentally demonstrate that
sufficient ultrafast refractive index changes can be obtained with
sufficient lateral homogeneity.

\begin{figure} [H]
  \includegraphics[width=\linewidth]{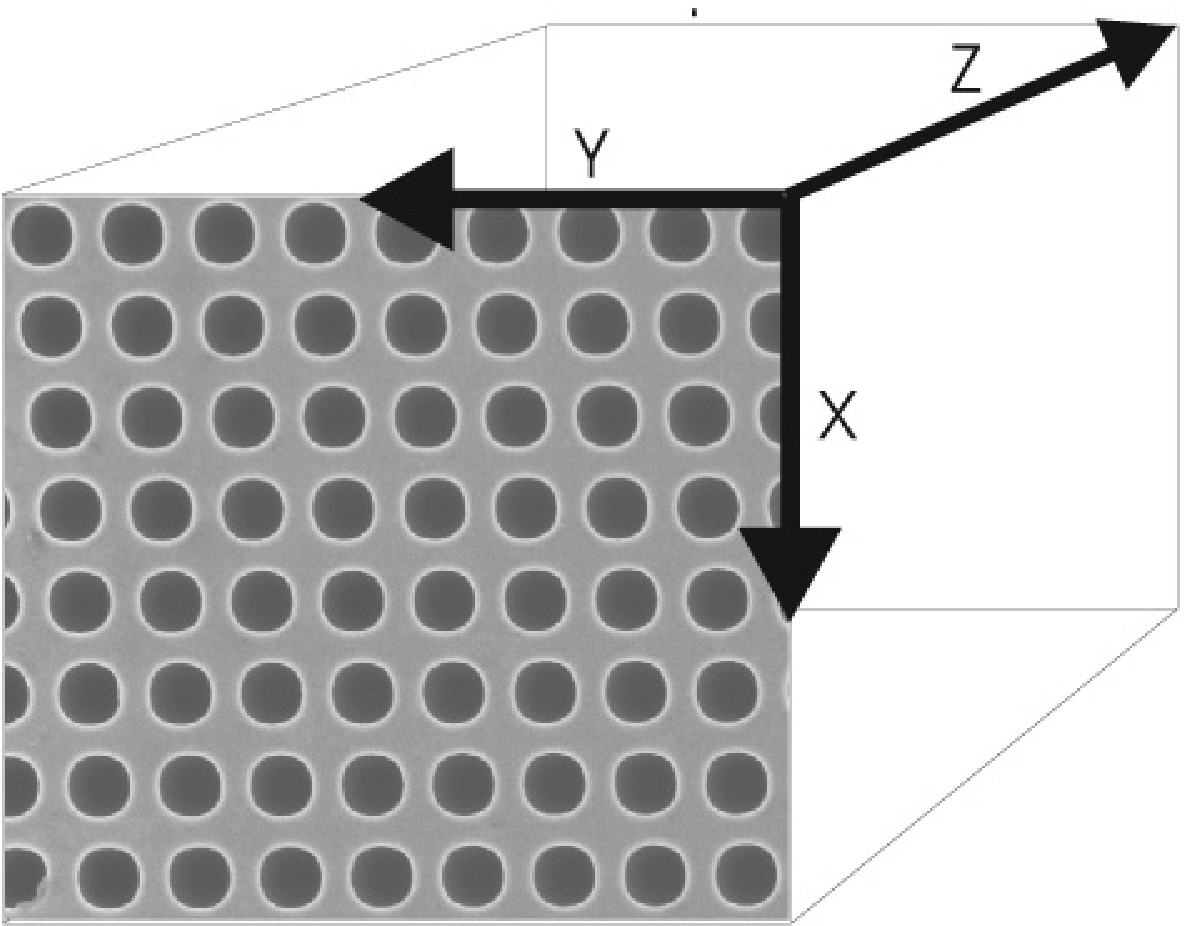}\\
  \caption{Schematic image of a sample: The z-axis is defined to be pointing into the sample,
  the x- and y-directions are the lateral directions. In the figure, the front face of the sample displays
  a SEM image of a dry etched Si 2D photonic crystal with a slightly rhomboid symmetry (lattice angle $85~\deg$).
  The lattice parameter $a$ is equal to $750$~nm. SEM courtesy of L.Woldering.}
  \label{fig:figure1}
\end{figure}

We concentrate on silicon, not only because of its wide
technological use, but also because it allows greater homogeneity
than GaAs that was discussed in Ref.~\onlinecite{Johnson02}.
Examples are given for light in the telecom band at
$\nu_{\rm{tele}}$/c~=~6450~ cm$^{-1}$ ($\lambda$~=~1550~nm), but
can easily be generalized to other frequencies. We set a minimum
volume of 5 unit cells cubed in which the change $\Delta n'$ must
remain within $10\%$ of its maximum value. This homogeneity
requirement holds for both the lateral directions x and y, as well
as for the z-direction defined in figure \ref{fig:figure1}. Even
in such a small crystal volume, the DOS already shows a
significant decrease for frequencies that lie in the band
gap.\cite{Kole} The typical unit cell size of silicon inverse opal
photonic crystals with a bandgap near $\nu_{\rm{tele}}$ is
$a$~=~1.2~$\rm{\mu}$m, therefore the homogeneously switched area
must extend at least 6~$\rm{\mu}$m in all three dimensions.

\section{Free carrier generation}

 In optical free carrier generation a pump pulse
is absorbed by a semiconductor sample, creating a free carrier
plasma with electron-hole density N$_{\rm{eh}}$. The generated
carrier plasma changes the dielectric function $\epsilon(\omega)$
of the sample by an amount $\Delta\epsilon_{\rm{eh}}(\omega)$. In
the case of silicon, the Drude model gives a excellent description
of the resulting $\epsilon(\omega)$ for densities below
$10^{22}$~cm$^{-3}$ (see Ref.~\onlinecite{Sokolowski00}).

\begin{equation}\label{eq:drudemodel}
    \epsilon(\omega)=\epsilon_B(\omega)+\Delta\epsilon_{\rm{eh}}(\omega)
    =\epsilon_B-\Big(\frac{\omega_p}{\omega}\Big)^2\frac{1}{1+i\frac{1}{\omega\tau_D}},
\end{equation}

where $\epsilon_B$ is the bulk dielectric constant, $\omega$
(=$2\pi\nu$) the frequency of the probe light, $m_e$ the electron
mass, $m_{opt}^*$ the optical effective mass of the carriers,
$\tau_D$ the Drude damping time,\cite{Siparam} and
$\omega_p$~=~$\sqrt{(N_{eh}e^2)/(\epsilon_0~ m_{opt}^* m_e)}$ the
plasma frequency. Under the condition that
$1/(\omega\tau_D)~\ll~1$, which is valid for silicon at
$\omega_{\rm{tele}}$, we can derive the following simplified
expression for the refractive index from
Eq.~(\ref{eq:drudemodel}):

\begin{equation}\label{eq:imagn}
    n'+in''= \sqrt{\epsilon(\omega)}=
    \sqrt{\epsilon_B-(\frac{\omega_p}{\omega})^2} + i\frac{\omega_p^2}{2\omega^3 \tau_D
    \sqrt{\epsilon_B-(\frac{\omega_p}{\omega})^2}}.
\end{equation}

For silicon with carrier densities below $10^{20}$~cm$^{-3}$, the
refractive index is linear with the carrier density within
$0.2\%$:

\begin{equation}\label{eq:linear_n_Neh}
     n'=\sqrt{\epsilon_B}-\frac{e^2}{2\sqrt{\epsilon_B}\epsilon_0 m_{opt}^* m_e
     \omega^2}N_{\rm{eh}}.
\end{equation}

Thus, the induced change in the refractive index $n'$ (via
$\epsilon(\omega)$) is completely determined by the optically
induced carrier density.

\begin{figure} [H]
  \includegraphics[width=\linewidth]{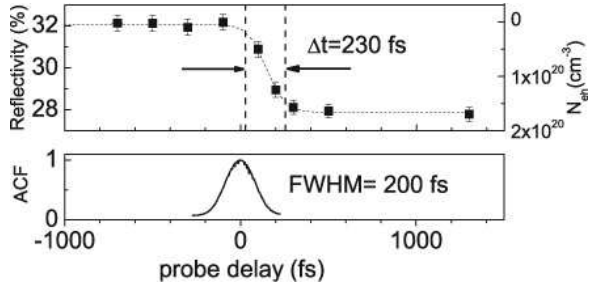}\\
  \caption{Time resolved reflectivity measurement on bulk Si,
 pumped at  $\nu_{\rm{pump}}$/c~=~12500~cm$^{-1}$, pulse energy E$_{\rm{pump}}$~=~2~$\rm{\mu}$J,
 pulse duration $\tau_{\rm{pump}}$~=~120~fs, r$_{\rm{pump}}$~=~70~$\rm{\mu}$m and peak intensity
115~GW/cm$^2$ (upper panel). The reflectivity of a probe with
$\nu_{\rm{probe}}$/c~=~7692~cm$^{-1}$,
r$_{\rm{probe}}$~=~20~$\rm{\mu}$m and $\tau_{\rm{probe}}$=120~fs
decreases from $32\%$ to $28\%$, corresponding to a calculated
carrier density N$_{\rm{eh}}$~=~$1.6\times 10^{20}$~cm$^{-3}$ at
the surface of the sample (see right-hand scale). The time
difference between $10\%$ and $90\%$ of the total change
  is 230~fs. The lower panel shows the intensity autocorrelation function
  (ACF) of the pump pulses. The full width half maximum (FWHM) of 200~fs
  corresponds to a pulse width of 140~fs FWHM.}
  \label{fig:figure2}
\end{figure}

An example of a carrier induced change of refractive index in bulk
silicon is given in Fig.~\ref{fig:figure2}. In this experiment, a
powerful ultrashort pump pulse was focussed to a spot with radius
r$_{\rm{pump}}$~=~70~$\rm{\mu}$m, resulting in a peak intensity at
the sample interface of I$_{\rm{0}}$~=~115~GWcm$^{-2}$. The
reflectivity of a weaker probe pulse with a smaller spot radius of
r$_{\rm{probe}}$~=~20~$\rm{\mu}$m was measured in the center of
the pumped spot at different time delays with respect to the pump
pulse. The scan in Fig.~\ref{fig:figure2} shows that the
reflectivity of the sample changes from $32\%$ to $28\%$. The
$10\%-90\%$ rise time is 230 fs, confirming an ultrafast change in
$n'$. From Fresnel's formula we find the refractive index change
to be more than $10\%$, corresponding to a large generated carrier
density of $1.6\times10^{20}$~cm$^{-3}$. This is about twice the
carrier density that is needed to obtain a change in $n'$ of $5\%$
for probe light at $\omega_{\rm{tele}}$. We have shown that
optical carrier generation can be used to change $n'$ by a large
amount on a sub picosecond timescale, meeting the first two of the
four main density of states switching requirements.

For applications such as optical modulators and waveguides, much
smaller changes in $n'$ are already sufficient, typically $\Delta
n'$~=~$10^{-4}$, see Ref.~\onlinecite{intel04}. This corresponds
to a carrier density of $10^{16}$~cm$^{-3}$. For such low carrier
densities, the required pump pulse energy at
$\nu_{\rm{pump}}$/c~=~12500~cm$^{-1}$ is on the order of several
nJ, allowing the use of diode lasers with repetition rates
exceeding 1 GHz as a pump source. Thus it seems that
carrier-induced optical switching may have much broader
applications beyond photonic crystals.

Next, we will discuss the induced absorption in switched
semiconductors, which is the third density of states switching
requirement. The carrier absorption length $\ell_{\rm{ca}}$ of the
excited carrier plasma is equal to

\begin{equation}\label{eq:abslength}
    \ell_{\rm{ca}}=\frac{1}{2k_0~n''(\omega)}
\end{equation}

where $k_0$~=~$\omega n'(\omega)/c$. After inserting the expressions
for $n'(\omega)$ and $n''(\omega)$ from Eq.~(\ref{eq:imagn}) we
get:

\begin{equation}\label{eq:abslength2}
    \ell_{\rm{ca}}=\bigg(\frac{\omega}{\omega_p}\bigg)^2 \tau_D c.
\end{equation}

We immediately see that the absorption length is inversely
proportional to the plasma frequency $\omega_p$ squared and thus
inversely proportional to the carrier density. In silicon, the
carrier absorption length  of probe light at $\omega_{\rm{tele}}$
is $\ell_{\rm{ca}}$~=~22~$\rm{\mu}$m for a carrier density of
$10^{20}$~cm$^{-3}$. In our analysis of 3D photonic crystals we
assume the crystals to be inverse opals with a typical filling
silicon filling fraction of $\Phi$~=~25$\%$. The carrier
absorption length in such crystals will then be approximately four
times larger than the bulk carrier absorption length, or
$\ell_{\rm{ca}}$~=~88~$\rm{\mu}$m. This result shows that in our
analysis, where the carrier density is less than
$10^{20}$~cm$^{-3}$, the carrier induced absorption for
$\omega_{\rm{tele}}$ remains very small. We have now shown that
the first three requirements for successful switching experiments
can be met for silicon at telecom frequencies. The remainder of
this paper will discuss the fourth requirement: the spatial
switching homogeneity.

\section{Optical properties of silicon}

\begin{figure} [H]
  \includegraphics[width=\linewidth]{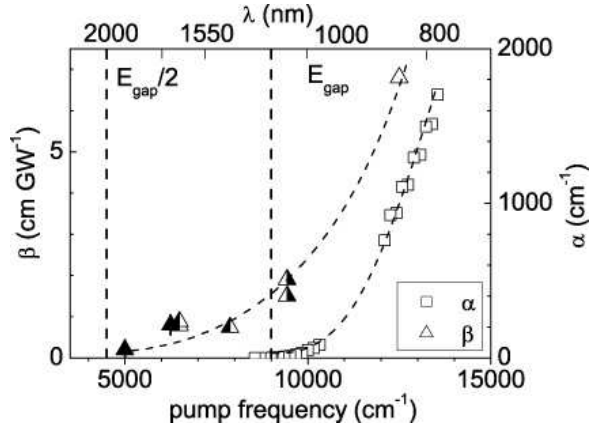}\\
  \caption{Absorption coefficients $\alpha$ and $\beta$ of Si versus pump
  frequency. The dashed vertical lines correspond to photon energies of
  E$_{\rm{gap}}$/2 and E$_{\rm{gap}}$. The open squares indicate linear absorption coefficients taken from
  Ref.~\onlinecite{Palik85}(right-hand scale). The solid triangles indicate the two-photon absorption
  coefficients that we have determined by z-scan measurements (left-hand
  scale). The left-filled triangles indicate the values from Ref.~\onlinecite{Dinu03}, the right-filled
  triangles are data from Ref.~\onlinecite{Rein73}, and the open triangle data from Ref.~\onlinecite{Sab02}}
  \label{fig:figure3}
\end{figure}

To analyze the switching homogeneity in the z direction, we
briefly consider how pump light is absorbed in semiconductors. At
low pump intensities I, the absorption of light in semiconductors
scales with the intensity: $\alpha$I, where the absorption
coefficient $\alpha$ tends to zero for photon energies
$\hbar\omega$ below the electronic bandgap energy E$_{\rm{gap}}$,
see Fig.~\ref{fig:figure3} for silicon. At high pumping
intensities, non linear two-photon absorption starts to play an
important role. For two-photon absorption, the absorption is
proportional to $\beta I^2$ where $\beta$ is the two-photon
absorption coefficient. This coefficient is expected to vanish for
$\hbar\omega$~$>$~E$_{\rm{gap}}$/2. Fig.~\ref{fig:figure3}
displays the frequency dependence of the linear and two-photon
absorption coefficients of silicon. The data was obtained from our
measurements and from Refs.\cite{Palik85,Dinu03,Rein73,Sab02}:
With our lasersystem,\cite{lasersystem}  z-scan
measurements\cite{She90} were performed to obtain the two-photon
absorption coefficient $\beta$ at two additional wavelengths. For
bulk silicon at $\nu_{\rm{pump}}$/c~=~5000~cm$^{-1}$, $\beta$ was
measured to be 0.20~$\pm$~0.05~cmGW$^{-1}$. At
$\nu_{\rm{pump}}$/c~=~6250~cm$^{-1}$, we obtained
$\beta$~=~0.80~$\pm$~0.1~cmGW$^{-1}$. The latter value is in
excellent correspondence with the value
$\beta$~=~0.88~$\pm$~0.13~cmGW$^{-1}$ at
$\nu_{\rm{pump}}~$/c=~6494~cm$^{-1}$ from
Ref.~\onlinecite{Dinu03}. Our measurements confirm that $\beta$
tends to zero for photon energies approaching E$_{\rm{gap}}$/2.
Both the linear and the nonlinear absorption coefficient can thus
be controlled by varying the pump frequency.

\section{Homogeneity of switched semiconductors}

\subsection{Homogeneity in the z-direction}

We now present a model that calculates the carrier density depth
profile N$_{\rm{eh}}$(z) caused by absorption of pump light. The
absorbed intensity is described by the nonlinear differential
equation

\begin{equation}\label{eq:absorption}
    \frac{dI(z)}{dz}=-[\alpha I(z)+\beta I^2(z)],
\end{equation}

which we have solved by implicit integration. The resulting
expression describing the intensity depth profile is

\begin{equation}\label{eq:depthprofile}
    I(z)=\frac{I_0e^{-\alpha z}}{1+(\beta I_0/\alpha)(1-e^{-\alpha
    z})},
\end{equation}

where I$_{\rm{0}}$ is the intensity at the interface. The
resulting carrier density profile N$_{\rm{eh}}$(z) is related to
the intensity depth profile I(z) as

\begin{equation}\label{eq:densityprofile}
    \rm{N_{eh}}(z)= \frac{I(z)\tau_{pump}}{\hbar \omega_{pump}}\Big[\alpha+\frac{1}{2}\beta
    I(z) \Big],
\end{equation}

where $\tau_{\rm{pump}}$ is the pump pulse duration. The factor
$1/2$ for two-photon absorption indicates that two photons must be
absorbed to generate one electron-hole pair. Substitution of
Eq.~(\ref{eq:depthprofile}) into Eq.~(\ref{eq:densityprofile})
allows us to calculate the carrier density profile for any given
combination of $\alpha$, $\beta$ and I$_{\rm{0}}$.

We now investigate the relation between the absorption
coefficients and pumping homogeneity. To quantify the homogeneity
of a switched sample, we first define the homogeneity length
$\ell_{\rm{hom}}$ within which the carrier density remains within
$10\%$ of its surface value:

\begin{equation}
 \label{eq:lhom}
 \ell_{\rm{hom}}\equiv0.1\times\bigg[\frac{1}{\rm{N_{eh}(z)}}\frac{\rm{dN_{eh}(z)}}{dz}\bigg]_{z=0}.
\end{equation}

\begin{figure} [H]
  \includegraphics[width=\linewidth]{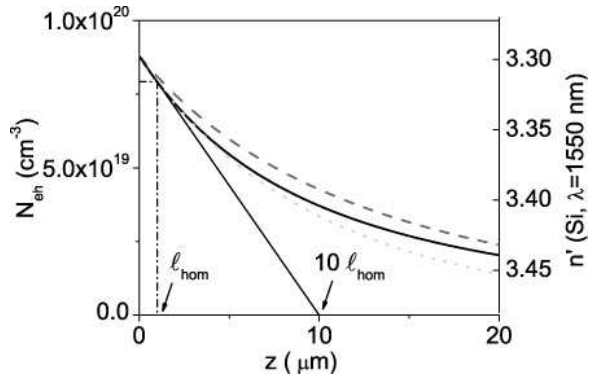}\\
  \caption{Calculated carrier density depth profile N$_{\rm{eh}}$(z) for constant
  $\beta$~=~2~cmGW$^{-1}$ and three different linear absorption coefficients:
  $\alpha$~=~0 (solid curve), $\alpha$~=~400~cm$^{-1}$ (dashed curve) and
  $\alpha$~=~800~cm$^{-1}$ (dotted curve). With N$_{\rm{eh}}$ at the interface
  kept constant at $0.9\times~10^{20}$~cm$^{-3}$, the necessary pump intensity
  I$_{\rm\breve{}{0}}$ is calculated, assuming that $\tau_{\rm{pump}}$~=~120 fs and r$_{\rm{pump}}$~=~75~$\rm{\mu}$m.
  The corresponding carrier density depth profile was obtained
  with Eq.~(\ref{eq:depthprofile}). The homogeneity length $\ell_{\rm{hom}}$ is shown
  for $\alpha$~=~0. The right-hand scale shows the resulting refractive index $n'$
  for $\nu$/c~=~6450~cm$^{-1}$ light in Si.}
    \label{fig:figure4}
\end{figure}

Because the homogeneity length is directly related to the maximum
gradient in the carrier depth profile, it is a helpful parameter
in quantifying the homogeneity of switched semiconductors. For
applications where switching homogeneity is important, the
homogeneity length should be much larger than the size of the
switched sample. To illustrate the homogeneity length, three
carrier density depth profiles are shown in Fig.~\ref{fig:figure4}
for which $\beta$ is kept constant at 2~cmGW$^{-1}$, and $\alpha$
is varied from zero (dotted curve) to 400~cm$^{-1}$ (dashed curve)
to 800~cm$^{-1}$ (solid curve). For each case, I$_{\rm{0}}$ is
chosen such that the carrier density reaches
$0.9\times10^{20}$~cm$^{-3}$ at the sample interface,
corresponding to a $5\%$ change in $n'$ at $\nu_{\rm{tele}}$ in
silicon. On the right y-axis the corresponding real part of the
refractive index $n'$ for silicon at $\nu_{\rm{tele}}$ is shown.
The pump frequency was assumed to be 5000~cm$^{-1}$. A closer look
at the three depth profiles in Fig.~\ref{fig:figure4} shows that
for $\alpha$~=~0~cm$^{-1}$, the homogeneity length is
$1.0~\rm{\mu}$m. As $\alpha$ increases to 400~cm$^{-1}$, the
homogeneity length increases to 1.2~$\rm{\mu}$m. As $\alpha$
increase further to 800~cm$^{-1}$, the homogeneity length
decreases again to $1.0~\rm{\mu}$m. The surprising occurrence of a
maximum in the homogeneity length can be explained with the aid of
Eq.~(\ref{eq:densityprofile}). For small $\alpha$
($\alpha~\ll~\rm{I_0}\beta$), the absorption is dominated by
two-photon absorption. If $\alpha$ increases, the pump intensity
I$_{\rm{0}}$ needed to obtain the surface carrier density
decreases, reducing the slope of the intensity profile at the
interface, determined by the exponent $-(\alpha+I_0\beta)$. This
leads to an increase in $\ell_{\rm{hom}}$. If $\alpha$ increases
further to the regime where the absorption is dominated by linear
absorption ($\alpha~\gg~\rm{I_0}\beta$), any further increase in
$\alpha$ will result in a decrease of the homogeneity length. In
the region between the two extremes, the homogeneity length
apparently attains a maximum value. This means that simply
choosing two-photon (or perhaps even higher-photon) absorption
over linear absorption is not always sufficient to ensure an
optimal homogeneity.

\begin{figure} [H]
  \includegraphics[width=\linewidth]{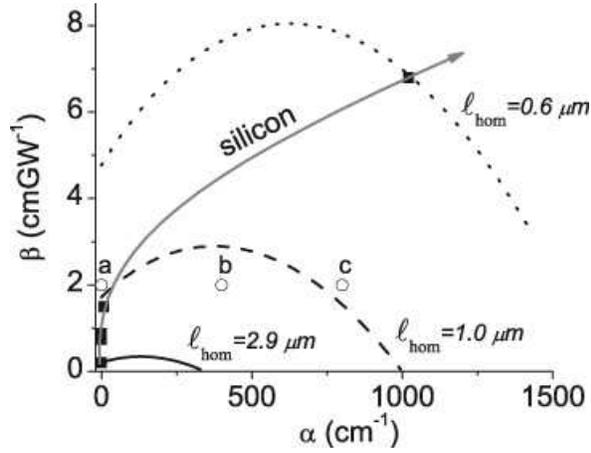}\\
  \caption{Homogeneity plot. Contours of constant $\ell_{\rm{hom}}$ are plotted
  in the ($\alpha,\beta$) plane, defining regions in which homogeneous switching can
  be achieved. The generated carrier density is kept constant at N$_{\rm{eh}}$(0)~=~$0.9\times10^{20}$~cm$^{-3}$,
  enough for a $5\%$ change in $n'$.  The solid curve corresponds to $\ell_{\rm{hom}}$~=~2.9~$\rm{\mu}$m, the dashed curve to
  $\ell_{\rm{hom}}$~=~1.0~$\rm{\mu}$m and the dotted curve to $\ell_{\rm{hom}}$~=~0.6~$\rm{\mu}$m for bulk silicon. The closed squares connected
  by the dotted arrow are linear and two-photon coefficients for Si obtained from
  Fig.~\ref{fig:figure2}. The open circles a, b and c correspond to the depth profiles N$_{\rm{eh}}$(z) plotted in
  Fig.~\ref{fig:figure4}. Pumping parameters: $\tau_{\rm{pump}}$~=~120 fs,
  r$_{\rm{pump}}$~=~75~$\rm{\mu}$m and $\nu_{\rm{pump}}$/c~=~5000~cm$^{-1}$.}
  \label{fig:figure5}
\end{figure}

We make a homogeneity plot to obtain further insight in the
influence of $\alpha$ and $\beta$ on the homogeneity. First we
choose a fixed electron density at the interface
N$_{\rm{eh}}$(0)~=~$0.9\times 10^{20}$~cm$^{-3}$. The
corresponding homogeneity length contours are then deduced from
our absorption model, and visualized in a plane spanned by linear
and two-photon absorption coefficients, see
Fig.~\ref{fig:figure5}. The absorption coefficients for silicon at
various frequencies taken from Fig.~\ref{fig:figure3} are also
plotted in the plane in Fig.~\ref{fig:figure5}. The three depth
profiles shown in Fig.~\ref{fig:figure4} correspond to positions
a, b, and c in Fig.~\ref{fig:figure5}. To obtain a certain minimum
homogeneity, the absorption coefficients must remain below the
curve corresponding to the particular minimum homogeneity. This
graph thus allows us to directly obtain the homogeneity length
that can be obtained for a semiconductor at a certain frequency.
The homogeneity plot demonstrates how pumping homogeneity can be
maximized by choosing the appropriate pump frequency. Generally,
smaller absorption coefficients lead to an increased homogeneity.
This increase comes at the price of a higher necessary pump
intensity I$_{\rm{0}}$. From Fig.~\ref{fig:figure5} we conclude
that the most homogeneous switch for silicon can be achieved for
$\nu_{\rm{pump}}$/c~=~5000~cm$^{-1}$. For this pump frequency, the
homogeneity length is 2.9~$\rm{\mu}$m. For comparison: if
$\nu_{\rm{pump}}$/c is equal to 12500~cm$^{-1}$, the homogeneity
length is only 0.6~$\rm{\mu}$m. The necessary pump intensity
remains below the maximum available pump energy of our laser
system I$_{\rm{max}}$. Our generalized homogeneity plot is valid
for all materials and allows us to find optimum switching
conditions. The choice for pumping frequencies which are low in
the two-photon absorption regime drastically increases the pumping
homogeneity.

\begin{figure} [H]
  \includegraphics[width=\linewidth]{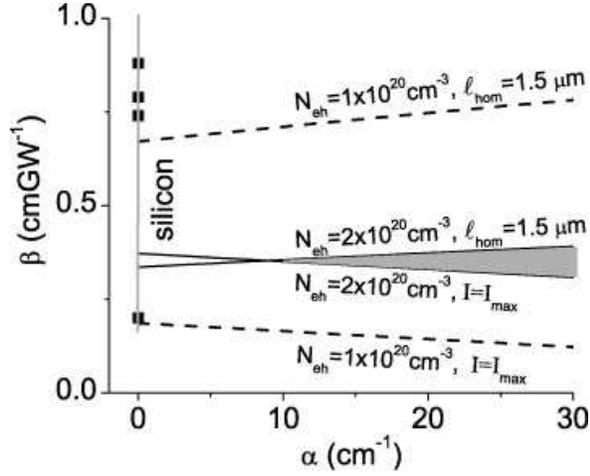}\\
  \caption{Carrier density contours for a particular bulk homogeneity
  chosen as $\ell_{\rm{hom}}$~=~1.5~$\rm{\mu}$m. The contours correspond to
  two different values for N$_{\rm{eh}}$(0): the dashed lines to
  N$_{\rm{eh}}$(0)~=~$1\times 10^{20}$~cm$^{-3}$, and the solid lines to
  N$_{\rm{eh}}$(0)~=~$2\times 10^{20}$~cm$^{-3}$. The upper one of each pair of
  curves corresponds to $\ell_{\rm{hom}}$~=~1.5~$\rm{\mu}$m. The lower curve of each
  pair indicates the minimum absorption coefficients for which the given
  N$_{\rm{eh}}$(0) can be obtained without exceeding the maximum
  intensity I$_{\rm{max}}$~=~1~TWcm$^{-2}$. The closed squares connected by the dotted curve line are the $\alpha$ and $\beta$
  coefficients for Si obtained from Fig.~\ref{fig:figure3}.  Pumping parameters: $\tau_{\rm{pump}}$~=~120 fs,
  r$_{\rm{pump}}$~=~75~$\rm{\mu}$m and $\nu_{\rm{pump}}$/c~=~5000~cm$^{-1}$.}
  \label{fig:figure6}
\end{figure}

We now make a plot that shows surface carrier density contours
(N$_{\rm{eh}}$(0)) corresponding to a particular fixed homogeneity
length. From such a plot the maximum homogeneous change in
refractive index can be derived. We choose a fixed homogeneity
length of $1.5~\rm{\mu}$m, since inside a typical photonic crystal
with $25\%$ filling fraction, the homogeneity length will be four
times larger ($6~\rm{\mu}$m) thus fulfilling our homogeneity
requirement. Fig.~\ref{fig:figure6} displays contours for two
different carrier densities
$_{\rm{eh}}$~=~$1\times10^{20}$~cm$^{-3}$ and
$2\times10^{20}$~cm$^{-3}$.

In Fig. \ref{fig:figure6}, the upper curve is the homogeneity
contour, and the lower one of each pair of curves indicates the
minimum absorption coefficients for which carrier density can be
generated given the maximum available intensity
I$_{\rm{max}}$~=~1~TWcm$^{-2}$.\cite{lasersystem} For feasible
switching experiments, the absorption coefficients must be in the
area to the right of the intersection of the two curves. As the
carrier density increases, the constant $\ell_{\rm{hom}}$ line
moves towards lower absorption coefficients, while the
I$_{\rm{max}}$ line moves towards higher values. As an example:
for a carrier density of $2\times 10^{20}$~cm$^{-3}$, homogeneous
switching can only be achieved for absorption coefficients within
the shaded area in Fig.~\ref{fig:figure6}. It is seen that this
area does not overlap with the trajectory of silicon parameters.
Therefore, this carrier density is not achievable given
$\ell_{\rm{hom}}$~=~$1.5~\rm{\mu}$m. With decreasing carrier
density, the range between the curves will overlap the silicon
parameter trajectory at some point. Such an intersection
determines the upper limit to the carrier density (given
$\ell_{\rm{hom}}$~=~$1.5~\rm{\mu}$m) as well as the pump frequency
that pertains to the relevant ($\alpha$,$\beta$)-point. For
silicon, this intersection is calculated to occur at
N$_{\rm{eh}}$~=~$1.9\times10^{20}$~cm$^{-3} $. From
Eq.~(\ref{eq:linear_n_Neh}) we obtain the corresponding maximum
homogeneous change in refractive index at $\nu_{\rm{tele}}$ to be
$11\%$.

The condition of a maximum intensity I$_{\rm{max}}$ can be relaxed
by choosing a smaller pump spot radius r$_{\rm{pump}}$. The
assumed pump radius of 75~$\rm{\mu}$m provides a lateral
homogeneity which is large compared to the homogeneity
requirement, therefore we could choose a smaller r$_{\rm{pump}}$
while maintaining sufficient lateral homogeneity. A higher pump
intensity would allow homogeneous switching experiments at even
lower absorption coefficients, allowing larger changes of the
refractive index. However, we must keep in mind that the carrier
absorption length is inversely proportional to the induced carrier
density, see Eq.~(\ref{eq:abslength2}). For a refractive index
change of $11\%$, we predict the carrier absorption length inside
a photonic crystal to drop to $\ell_{\rm{ca}}$~=~38~$\rm{\mu}$m,
which may be sufficient to meet our third requirement for
switching experiments in photonic crystals. At higher carrier
densities however, the carrier absorption length will become too
small to meet this requirement.

A similar analysis can be done for other semiconductor materials.
The carrier density needed for a $5\%$ change in $n'$ will depend
on material properties such as the optical effective mass of the
carriers and the Drude damping time $\tau_D$ (see
Eq.~(\ref{eq:drudemodel})). We briefly discuss switching of GaAs
at a frequency $\nu$/c~=~9430~cm$^{-1}$ proposed in
Ref.~\onlinecite{Johnson02}. For GaAs the optical effective mass
$m_{opt}^*$~=~0.06 is 2.5 times smaller than for
silicon.\cite{dargys} Therefore, the carrier density in GaAs is
2.5 times lower than what is required for the same refractive
index change in silicon. We find that the area increases in which
the homogeneity requirement can be met. However, the two-photon
absorption coefficient for GaAs at $\nu$/c~=~9430~cm$^{-1}$ is
26~cmGW$^{-1}$, far above the homogeneous switching
area.\cite{dargys} This immediately shows, that GaAs at this pump
frequency cannot be used in homogeneous switching experiments.
From this brief analysis we learn that the pumping frequencies in
GaAs should be reduced, to where the two-photon absorption
coefficient is much lower value.

\subsection{Lateral homogeneity}

We now consider the homogeneity in the lateral (x,y) directions.
We have measured the relative change in probe reflectivity from a
bulk silicon sample pumped in the two-photon absorption regime at
$\nu_{\rm{pump}}$/c~=~6250~cm$^{-1}$. For this frequency,
$\beta$~=~0.8~$\pm$~0.1~cmGW$^{-1}$ and $\alpha$~=~0. The probe
delay is fixed at 10~ps to avoid transient effects. The pump focus
position is shifted with respect to the probe focus by adjusting
the pump mirror with a micrometer drive. The pump energy
E$_{\rm{pump}}$ is 10.8~$\rm{\mu}$J on a focus with
r$_{\rm{pump}}$~=~80~$\rm{\mu}$m. The probe frequency
$\nu_{\rm{probe}}$/c is 7692~cm$^{-1}$, with a focus size of
$r_{\rm{probe}}$~=~25~$\rm{\mu}$m. The dashed line in the graph is
drawn to indicate the width of the pump intensity distribution.
The measured relative reflectivity shows a minimum at the center
of the pump distribution, and decreases away from the center. The
reflectivity data was fitted by a Gaussian curve (solid curve).
The radius of this measured reflectivity minimum is
34~$\pm$~5~$\rm{\mu}$m, which is considerably smaller than the
r$_{\rm{pump}}$. To obtain the lateral homogeneity length, we have
determined the maximum lateral distance by the pump focus center
for which $\Delta$R/R remains within $10\%$ of its maximum value
(dotted lines). This distance turns out to be 16~$\rm{\mu}$m in
both directions, corresponding to a homogeneity length of
32~$\rm{\mu}$m. This illustrates that for a sample pumped by a
pump beam with r$_{\rm{pump}}$~=~80~$\rm{\mu}$m, the lateral
homogeneity length is much better than the 6~$\rm{\mu}$m which we
required for switching of the density of states.

\section{Switching homogeneity in real photonic crystals}

The above analysis was done for bulk semiconductor samples. To
obtain the homogeneity length in photonic crystals, the bulk
absorption length is divided by the semiconductor filling fraction
$\Phi$. In case of a $5\%$ switch of the refractive index in a
silicon photonic crystal with $\Phi$~=~$25\%$ at
$\nu_{\rm{pump}}$/c~=~5000~cm$^{-1}$, the homogeneity length thus
increases from 2.9~$\rm{\mu}$m to 11.6~$\rm{\mu}$m, which is twice
the homogeneity requirement of 6~$\rm{\mu}$m. For a higher pump
frequency of 12500~cm$^{-1}$, we find a homogeneity length of
2.3~$\rm{\mu}$m, which is too low. This illustrates that the
homogeneity required for switching of the density of states in
silicon photonic crystals can only be performed at two-photon
absorption frequencies.

In the analysis so far, the extinction of pump light due to random
scattering inside the photonic crystals was neglected. We now
discuss how to incorporate inevitable scattering in photonic
samples. Scattering is quantified by the mean free path
$\ell_{\rm{mfp}}$: the characteristic length over which a coherent
beam becomes diffuse. The homogeneity length of light inside a
photonic crystal is related to $\ell_{\rm{mfp}}$ and
$\ell_{\rm{abs}}$ as

\begin{equation}\label{eq:lext}
\ell_{\rm{hom}}=0.1\times\bigg[\frac{1}{\ell_{\rm{abs}}}+\frac{1}{\ell_{\rm{mfp}}}\bigg]^{-1}.
\end{equation}

In the limit of weak scattering, where
$\ell_{\rm{mfp}}$~$\gg$~$\ell_{\rm{abs}}$, Eq.\ref{eq:lext}
reduces to $\ell_{\rm{hom}}$~=~0.1$\times \ell_{\rm{abs}}$. In the
limit of strongly scatter, where
$\ell_{\rm{mfp}}$~$\ll$~$\ell_{\rm{abs}}$, the homogeneity length
becomes $\ell+{\rm{hom}}$~=~0.1$\times \ell_{\rm{mfp}}$. As
opposed to the adverse effect of scattering on the homogeneity in
the z-direction, scattering will generally be favorable for
lateral homogeneity, as pump light which is removed from the
coherent pump beam is be scattered laterally.

Recently, our group has developed a quantitative model of the mean
free path inside photonic crystals.\cite{Koend04} One of the main
results of the analysis is that the mean free path for pump light
decreases with pump frequency squared $\omega_{\rm{pump}}^{-2}$.
The absolute value of the mean free path depends not only on
frequency, but also on many properties of the crystal such as the
unit cell size, the amount of disorder and the refractive index
contrast. As an example, we apply the model to a silicon inverse
opal photonic crystal with a lattice parameter of $a$~=~1240~nm,
corresponding to a bandgap frequency near $\omega_{\rm{tele}}$. We
assume combined size poly dispersity and lattice displacements of
$2\%$ (which is beyond the current state of the art). From the
model of Ref.~\onlinecite{Koend04}, we obtain a mean free path of
5.3~unit cells for pump light at
$\nu_{\rm{pump}}$/c~=~12500~cm$^{-1}$. The corresponding
homogeneity length, dominated by $\ell_{\rm{mfp}}$, would thus be
0.5$~$unit cells, well below our homogeneity requirement of 5~unit
cells. For a 2.5 times lower pump frequency of 5000~cm$^{-1}$, the
model predicts a $(2.5)^2$ times larger mean free path of
$\ell_{\rm{mfp}}$~=~33.3~unit cells inside the crystal. Together
with the earlier obtained absorption length of
$\ell_{\rm{abs}}$~=~116~$\rm{\mu}$m~(=~93~unit cells), we obtain a
homogeneity length of $[(1/93+1/33)^{-1}]/10$~=~2.4~unit cells.
This result points out that even at pump frequencies near the
two-photon absorption edge of silicon, the homogeneity requirement
of $5$~unit cells can not be met in silicon inverse opals with a
lattice parameter of $a$~=~1240~nm.

The scattering model predicts that decreasing the unit cell size
will result in higher homogeneity, as this will reduce the
relative pump frequency. Therefore, we consider silicon inverse
opals with a reduced lattice parameter
 of $a$~=~900~nm, corresponding to a bandgap near 8900~cm$^{-1}$, just below
 the absorption edge of silicon. The scattering model predicts a larger mean free
path for pump light with a frequency of 5000~cm$^{-1}$ of
$\ell_{\rm{mfp}}$~=~63~unit cells. Using the absorption length of
116~$\rm{\mu}$m~(=~116~unit cells), we obtain homogeneity length
of: $\ell_{\rm{hom}}$~=~[$(1/116+1/63)^{-1}$]/10~=~4.1~unit cells,
close to our homogeneity requirement.

Major improvements in switching homogeneity can be made by
studying diamond-like structures,\cite{ho90,hillebrand03} as the
lattice parameter can be as low as 600~nm for a bandgap near
$\omega_{\rm{tele}}$ in such crystals. In such structures, the
scattering model remains unchanged, apart from a constant
prefactor that depends on the shape of the unit cell. We predict
that reduced lattice parameters will decrease the relative pump
frequency sufficiently to allow homogeneous photonic density of
states switching experiments.

\section{Conclusions}
We have discussed four important requirements for free-carrier
induced optical changes in semiconductors, with emphasis on
density of states switching in photonic crystals: the amplitude of
change in refractive index, the timescale on which the switch
takes place, the induced absorption, and the homogeneity of the
induced change. We have demonstrated that the first two
requirements can be met: a $10\%$ change in $n'$ in bulk silicon
observed within 230~fs. We have also experimentally demonstrated
sufficient lateral homogeneity in a bulk silicon. Furthermore, we
have measured two-photon absorption coefficients bulk silicon. The
induced absorption was deduced to be low for carrier densities
below $10^{20}$~cm$^{-3}$.

We have discussed a non linear absorption model to describe the
spatial homogeneity of optically generated electron-hole plasmas
in semiconductors. We have introduced a homogeneity plot, which
directly relates linear and two-photon absorption coefficients to
the maximum homogeneity that can be achieved for any
semiconductor. From such a plot, we conclude that for density of
states switching in silicon photonic crystals, the optimum carrier
density is about $10^{20}$~cm$^{-3}$. To obtain the required
homogeneity, the absorption coefficients must be minimized, by a
judicious choice of pump frequency. Due to peak intensity
limitations the lowest pump frequency that can be chosen is around
5000~cm$^{-1}$.

We have discussed the effect of scattering in photonic crystals on
the pumping homogeneity. The homogeneity of switched photonic
crystals turns out to be limited by scattering. We conclude that
the homogeneity condition can barely be met in silicon inverse
opals. We predict that in diamond structures the relative pump
frequency will be small enough to allow homogeneous density of
states switching experiments.

Finally, we have briefly discussed the ramifications for
waveguides and modulators and we conclude that free-carrier
switching is also useful for applications outside photonic
crystal.

\section{Acknowledgements}
 The authors wish to thank Irwan Setija
and Rutger Voets (ASML) for deep UV lithography, Meint de Boer
(TST) and Willem Tjerkstra for dry etching, and Leon Woldering for
sample preparation. We also thank Allard Mosk and Ad Lagendijk for
fruitful discussions and Martijn Wubs for deriving Eq.~(\ref{eq:depthprofile}). This work is part of the research program
of the "Stichting voor Fundamenteel Onderzoek der Materie" (FOM),
which is supported by the "Nederlandse Organisatie voor
Wetenschappelijk Onderzoek" (NWO).

\section{figures}

\begin{figure} [H]
  \includegraphics[width=\linewidth]{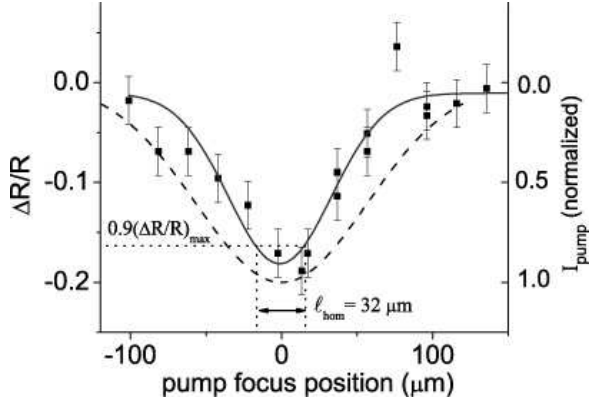}\\
  \caption{ Relative probe reflectivity measurement on a bulk
  Si sample for different lateral positions of the probe focus with respect to the
  probe focus. The probe delay is fixed at 10~ps. Pumping parameters:
  $\tau_{\rm{pump}}$~=~120~fs, $\nu_{\rm{pump}}$/c~=~6250~cm$^{-1}$, E$_{\rm{pump}}$~=~10.8~$\rm{\mu}$J,
  r$_{\rm{pump}}$~=~80~$\rm{\mu}$m, $\tau_{\rm{pump}}$~=~120~fs, $\nu_{\rm{probe}}$/c~=~7692~cm$^{-1}$,
  r$_{\rm{probe}}$~=~25~$\rm{\mu}$m. The squares are the measured datapoints, fitted with a Gaussian curve
  width of 68~$\pm$~10~$\rm{\mu}$m (solid curve). The horizontal dotted line indicates the
  level where the change in reflectivity has decreased by $10\%$. From the vertical dotted
  lines, which indicate the intersections of the $90\%$ line with the Gaussian fit of the data,
  we obtain a lateral homogeneity length of 32~$\rm{\mu}$m, illustrating the excellent lateral homogeneity.
  The dashed curve indicates the measured width of the pump focus.}
  \label{fig:figure7}
\end{figure}


\begin{thebibliography}{}
\bibitem{Siparam}
For silicon, the value of $\tau_D$ is $10^{-13}$, and the value
for $m_{opt}^*$ is 0.15, see Ref.~\onlinecite{Sokolowski00}.
\bibitem{lasersystem}
{ Our laser system provides high power pulses at two independently
tunable frequencies. The setup consists of a regeneratively
amplified Ti:Saf fs laser system (Spectra Physics Hurricane) which
is used to pump two Topas optical parametric amplifiers (OPA). The
OPA's have a continuously tunable output frequency between 3850
and 21050~cm$^{-1}$ and produce 120~fs pulses (for an
autocorrelation graph see Fig.~\ref{fig:figure2}, lower panel)
with a pulse energy $E_{pulse}$ of at least 20~$\rm{\mu}$J over
the whole frequency range. Given a focal radius of 75~$\rm{\mu}$m
the maximum intensity I$_{\rm{max}}$ is at least 1~TWcm$^{-2}$}
\end{thebibliography}
\end{document}